\documentclass{jfm_arxiv}

\usepackage{commath}
\usepackage[hidelinks]{hyperref}
\usepackage[nameinlink,noabbrev]{cleveref}%
\crefname{figure}{figure}{figures}
\crefname{equation}{equation}{equations}
\usepackage{graphicx}
\usepackage[numbers]{natbib}
\usepackage[locale=UK,
number-mode=math,
unit-mode=text,
per-mode=symbol,
number-unit-product=\ ,
retain-zero-exponent=false]{siunitx}[=v2]
\DeclareSIUnit{\pixel}{px}
\DeclareSIUnit{\fps}{fps}

\newcommand{\kindex}[2]{\ensuremath{{#1}_{\scalebox{0.5}{#2}}}}
\newcommand{\ksuper}[2]{\ensuremath{{#1}^{\scalebox{0.6}{#2}}}}
\newcommand{\ksubsuper}[3]{\ensuremath{{#1}_{\scalebox{0.5}{#2}}^{\scalebox{0.6}{#3}}}}
\newcommand{\subf}[1]{(\textit{#1})}

\crefformat{equation}{\eqA equation~\eqB #2#1#3)}
\newcommand{\eqA}{}
\newcommand{\eqB}{(}
\DeclareRobustCommand{\pcref}[1]{%
	\begingroup
	\renewcommand{\eqA}{(}\renewcommand{\eqB}{}%
	\cref{#1}%
	\endgroup
}

\graphicspath{{../figurematter/latex/figs/}}
\usepackage{todonotes,lipsum}
\usepackage{paralist,palatino,microtype,pifont,ifplatform}
\usepackage{xfrac,amssymb,bm,nicefrac,physics,floatrow}

\begin{document}
\title{Estimating the non-dimensional energy of vortex rings by modelling their roll-up}

\author{Guillaume de Guyon, Karen Mulleners$^{*}$}
\affiliation{Institute of mechanical engineering, \'Ecole polytechnique f\'ed\'erale de Lausanne, Switzerland\\
$^*$ Corresponding author: karen.mulleners@epfl.ch
}

\maketitle
\begin{abstract}
The non-dimensional energy of starting vortex rings typically converges to values around \num{0.33} when they are created by a cylinder piston or a bluff body translating at a constant speed. To explore the limits of the universality of this value and to analyse the variations that occur outside of those limits, we present an alternative approach to the slug-flow model to predict the non-dimensional energy of a vortex ring. Our approach is based on the self-similar vortex sheet roll-up described by \cite{Pullin1978}. We derive the vorticity distribution for the vortex core resulting from a spiralling shear layer roll-up and compute the associated non-dimensional energy. To  demonstrate the validity of our model for vortex rings generated through circular nozzles and in the wake of disks, we consider different velocity profiles of the vortex generator that follow a power law with a variable time exponent $m$. Higher values of $m$ indicate a more uniform vorticity distribution. For a constant velocity ($m=0$), our model yields a non-dimensional energy of $\ksuper{E}{*}=0.33$.
For a constant acceleration ($m=1$), we find $\ksuper{E}{*}=0.19$.
The limiting value $m \rightarrow \infty$ corresponds to a uniform vorticity distribution and leads to $\ksuper{E}{*}=0.16$, which is close to values found in literature for a Hill's spherical vortex. The radial diffusion of the vorticity within the vortex core results in the decrease of the non-dimensional energy.
For a constant velocity, we obtain realistic vorticity distributions by radially diffusing the vorticity distribution of the Pullin spiral and predict a decrease of the non-dimensional energy from \num{0.33} to \num{0.28}, in accordance with experimental results. Our proposed model offers a practical alternative to the existing slug flow model to predict the minimum non-dimensional energy of a vortex ring. The model is applicable to piston-generated and wake vortex rings and only requires the kinematics of the vortex generator as input.
\end{abstract}

\section{Introduction}
The vorticity distribution within a vortex ring can be quantified by its non-dimensional energy $\ksuper{E}{*}=E/\sqrt{I \Gamma^3}$, where $E$ and $I$ are the energy and impulse per unit of mass, and $\Gamma$ the circulation of the vortex.
Vortices with a uniform vorticity distribution, such as Hill's spherical vortex \citep{Hill1894a}, can have a non-dimensional energy as low as $\ksuper{E}{*}=0.16$.
Vortices with a concentrated core of vorticity reach higher non-dimensional energy \citep{Friedman1981}.
Vortex rings are often produced by ejecting a volume of fluid through a circular nozzle.
If the ejection velocity is constant, the non-dimensional energy of the generated vortex converges to values ranging from \numrange{0.27}{0.35} \citep{Gharib1998,Mohseni2001,Limbourg2021a,Danaila2008}.
The non-dimensional energy of vortex rings that are produced in the wake of disks or cones translating at a constant speed also converges towards $\ksuper{E}{*} \approx 0.3$ \citep{Yang2012,DeGuyon2021}.
The main topological difference between the formation of propulsive vortex rings through a circular nozzle or drag vortex rings behind a translating object is the location of the feeding shear layer with respect to the main axis of symmetry.
The shear layer that feeds propulsive vortices is located close to the axis of symmetry and propulsive vortices are basically fed from the inside.
Drag vortices are fed from the outside as these vortices form in the wake of bluff bodies and the shear layer is located at the edge of the body.
Despite this fundamental difference in the feeding of propulsive and drag vortices, they reach similar non-dimensional energy values and vorticity distributions.

The main factors that influence the vorticity distribution of vortex rings are the velocity profile in the nozzle and the diffusive effects in the vortex.
The results cited above were obtained for Reynolds number above \num{2000}.
For these Reynolds numbers, the flow in the nozzle is considered turbulent and the velocity profile across the nozzle diameter is rather uniform.
The more uniform the velocity profile across the outlet, the thinner the shear layer between the nozzle jet and the outer fluid.
This typically leads to vortex rings with a concentrated core of vorticity and a non-dimensional energy $\ksuper{E}{*} >0.25$.
For laminar flows, the velocity profile is parabolic and the non-dimensional energy decreases below \num{0.2} \citep{Rosenfeld1998,Kaplanski2012}.

A few convective times after the emergence of the vortex ring, its self-induced velocity exceeds the velocity of the shear layer that feeds it and the vortex separates or pinches off~\citep{Shusser2000}.
After vortex pinch-off, the circulation ceases to increase and the vorticity distribution remains constant at first, before diffusion becomes effective.
The non-dimensional time or stroke ratio at which the vortex separates is referred to as the vortex formation number.
The formation number refers to a limiting formation time above which an isolated vortex ring can not be formed.
The formation number of a vortex ring can be altered by progressively varying the fluid velocity at the exit of the nozzle \citep{Mohseni2001,Dabiri2005a,Shusser2006}.
By increasing the nozzle velocity in time, the shear layer velocity increases and remains higher than the vortex self-induced velocity for a longer time, increasing its formation number.
During the longer formation process, the circulation is distributed more uniformly within the vortex core region, yielding a lower non-dimensional energy.
Higher formation numbers are thus associated with lower values of \ksuper{E}{*} and vice-versa.
With a velocity programme that increases linearly in time, a piston-cylinder apparatus can create vortex rings with a non-dimensional energy $\ksuper{E}{*}$ between \num{0.17} and \num{0.22} \citep{Mohseni2001,Dabiri2005a}.

The most commonly used procedure to estimate the vortex formation number relying on the piston temporal velocity profile is based on the slug-flow model \citep{Gharib1998b,Shusser1999,Linden2001,Krieg2021,Limbourg2021b}.
The slug-flow model predicts the invariants of the fluid injected in the system based on the nozzle geometry and the piston velocity.
It assumes that the bulk of fluid ejected from the nozzle has a spatially uniform velocity profile.
Based on the predicted values of the circulation, impulse, and energy of the ejected fluid from the slug-flow model, we can calculate the translation velocity $U$ of a generated vortex ring using the thin core approximation \citep{Saffman1975}:
\begin{equation}
U = \frac{E}{2I} + \frac{3}{8} \sqrt{\frac{\Gamma^3}{\pi I}}\quad.
\end{equation}
The formation number can now be identified as the formation time at which the vortex translational velocity exceeds the velocity of the shear layer that feeds it.
This is a purely kinematic argument for vortex pinch-off.
It requires an accurate estimation of the shear layer velocity.
A raw approximation is to consider the shear layer velocity to be half the one of the piston, which leads to slightly underestimated formation number of \num{3}.
More accurate approximation were performed, based on the spatial velocity profile at the nozzle exit and the contraction ratio of the starting jet \citep{Weigand1997,Limbourg2021b}.
This procedure for estimating the formation number is know as the asymptotic matching procedure and was recently reviewed and extended for application to orifice-generated vortices by \citet{Limbourg2021c}.
The kinematic argument for vortex pinch-off is only suitable for propulsive vortices, which move away from the region where they are created.
Vortex rings that form in the wake of a bluff body have a self-induced velocity that allows them to follow the body even beyond their formation number \citep{DeGuyon2021}.

To identify the formation number and the minimal value of the non-dimensional energy for both propulsive and drag vortices, we revert to the energetic argument for vortex pinch-off.
According to the Kelvin-Benjamin variational principle, we can identify the formation number as the formation time at which the non-dimensional energy of the fluid injected in the system falls below the non-dimensional energy of the vortex ring \citep{Benjamin,Gharib1998}.
The non-dimensional energy of the fluid injected to create propulsive vortices is again estimated using the slug-flow model.
For a piston moving at a constant velocity, \citet{Shusser2000} found a value of $\ksuper{E}{*}=0.31$ at vortex pinch-off.
The same authors extended their analysis to situations where the temporal evolution of the piston velocity follows a power law \citep{Shusser1999}.
When the piston velocity profile is given by $u(t) \propto t^m$, with $m\geq 0$, vortices are generated with a non-dimensional energy of:
\begin{equation}\label{shuss}
\ksuper{E}{*}=\frac{3}{4 \sqrt{\pi}} \frac{2m+1}{4m+1}\quad.
\end{equation}
The results of the slug-flow model for \ksuper{E}{*} are in good agreement with experimental data of propulsive vortex rings \citep{Shusser1999}.
But, the slug-flow model is not able to describe the non-dimensional energy of the fluid injected in vortices rings forming in the wake of bluff bodies, as the volume and the velocity of the fluid injected in the drag vortices are not known a priori.\\

Here, we propose a different approach to estimate the minimum non-dimensional energy of vortex rings.
We specifically build upon the work of \cite{Pullin1978}.
\citeauthor{Pullin1978} studied the roll-up of a shear layer that is created at the tip of a semi-infinite object that is translated from rest by different velocity profiles $u(t) \propto t^m$.
By modelling the self-similar roll-up of the vortex sheet, \citeauthor{Pullin1978} derived the temporal growth of the vortex circulation.
For $m=0$, the circulation of the vortex grows with $t^{\nicefrac{1}{3}}$ and its radius with $t^{\nicefrac{2}{3}}$.
Although this model was developed for two-dimensional flows, it matches experimental observations on vortex rings \citep{Pullin1979,Maxworthy_1977}.
\citet{Nitsche1994} present numerical simulations of the roll-up of a shear layer into a vortex ring and confirm that \citeauthor{Pullin1978}'s asymptotic analysis can indeed be extended to vortex rings and yield correct predictions of the vortex radius.
To predict the resulting non-dimensional energy of rings vortices, we adopt Pullin's model to derive the vorticity distribution within the vortex core.
We demonstrate that the current approach is valid for vortex rings generated through circular nozzles and in the wake of disks.

\section{Pullin's spiral and its integral quantities}
\begin{figure}
	\includegraphics{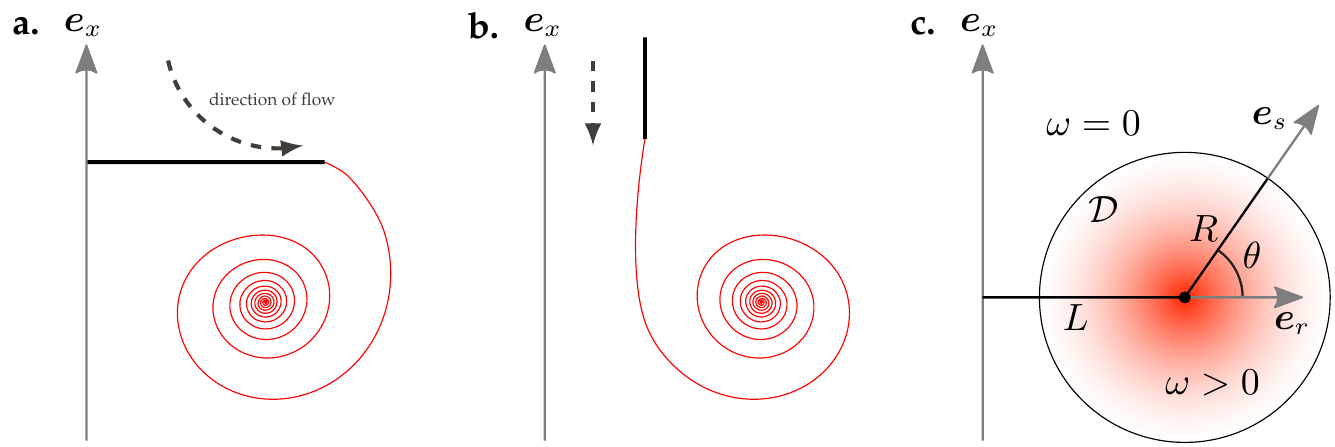}
	\caption[Shear layer roll-up]{Representation of vortex rings by the roll-up of a shear layer in cylindrical coordinates $(x,r)$.
		Vortex rings forming \subf{a} behind a disk, or \subf{b} at the exit of a circular nozzle.
		\subf{c} Sketch and definition of the parameters used to model the vortex ring.}
	\label{fig:Pull1}
\end{figure}
The roll-up of a shear layer past translating or rotating semi-infinite flat plates exhibits a self-similar behaviour \citep{Kaden, Pullin1978, Lepage2005, Xu2014, Pullin2020, Francescangeli.2021}.
Here, we focus our attention on parallel shear layers emanating from a piston-cylinder vortex generator or in the wake of an a translating disk.
The main parameter governing the roll-up is the velocity profile of the piston or of the disk, given by $u(t) \propto t^m$.
If $m=0$, the object moves with a constant speed.
If $m=1$, the object moves with a constant acceleration.
The vortex behind a disk (\cref{fig:Pull1}a) or at the exit of the circular nozzle (\cref{fig:Pull1}b) both originate from the roll-up of a shear layer into a spiral.
The drag vortex behind the disk is fed by the shear layer from the outside, whereas the propulsive vortex at the nozzle exit is fed by the shear layer from the inside.
The vortex core is characterised in a local polar coordinate system $(s,\theta)$, centred on the spiral point (\cref{fig:Pull1}c).
The vortex core coordinates are related to the original cylindrical coordinates $(x,r)$ by:
\begin{equation}
x=s\, R\, \sin \theta \quad \text{and} \quad r=L+s\,R\,\cos \theta\quad,
\end{equation}
with $L$ the distance from the centre of the spiral to the $x$-axis and $R$ the radius of the vortex core.
The coordinate $s$ is the radial coordinate relative to the core radius $R$.
\citeauthor{Pullin1978} derived an asymptotic solution for the circulation within a circular region of radius $s$ centred on the spiral:
\begin{equation}
\Gamma\left(s\right) = \dfrac{2 \pi \kindex{\omega}{0} R^2}{p} s^p\quad,
\label{eq:Pull1}
\end{equation}
with
\begin{equation}
p=2-\dfrac{3}{2(1+m)} \quad,
\label{eq:Pull1}
\end{equation}
and $\kindex{\omega}{0}$ a reference vorticity.
The circulation follows a radial power law distribution with an exponent $p$ that can range between \num{0.5} and \num{2}.
The vorticity distribution corresponding to this circulation is given by:
\begin{equation}
\omega(s)=\begin{cases}
\kindex{\omega}{0} s^{p-2} ,& \text{if } s\leq 1\\
0,              & \text{if } s> 1\quad.
\end{cases}
\label{eq:Pull2}
\end{equation}
We now compute the non-dimensional energy associated with this distribution.
The non-dimensional energy is defined as
\begin{equation}
\ksuper{E}{*}=\dfrac{E}{\sqrt{I\Gamma^3}} \quad,
\end{equation}
with
\begin{equation}
E= \pi \iint\displaylimits_{\mathcal{D}}\psi \omega \qq{,} I = \pi \iint\displaylimits_{\mathcal{D}} \omega r^2 \qq{,} \Gamma = \iint\displaylimits_{\mathcal{D}} \omega\quad.
\label{eq:ints}
\end{equation}
Here, $\mathcal{D}$ is the disk of radius $R$ centred on the vortex core (\cref{fig:Pull1}c).
The integration of the circulation and impulse associated with the vorticity distribution (\ref{eq:Pull1}) leads to
\begin{equation}\label{eq:IG}
	\Gamma = \dfrac{2 \pi \kindex{\omega}{0} R^2}{p} \qq{and}
	I=\dfrac{\kindex{\omega}{0} \pi^2 R^4}{p}\left(\dfrac{2}{\varepsilon^2}+\dfrac{p}{p+2} \right).
\end{equation}
The energy is expressed as a function of the stream function $\psi$:
\begin{equation}
\psi(x,r) = \iint\displaylimits_{\mathcal{D}}G(x,r,x',r')\,\omega(x',r')\dd{x'} \dd{r'} \quad,
\end{equation}
with $G$ the Green function
\begin{equation}
G(x,r,x',r')=\dfrac{\sqrt{rr'}}{2 \pi}\left[\left(\dfrac{2}{k}-k\right)K(k)-\dfrac{2}{k}E(k)\right]\quad,
\end{equation}
and
\begin{equation}
k^2=\dfrac{4rr'}{\left(x-x'\right)^2+\left(r+r'\right)^2} \quad .
\end{equation}
The relative vortex core radius is defined as $\varepsilon=R/L$.
\citeauthor{Pullin1978}'s model was initially derived for the roll-up of a vortex sheet that develops near the edge of a semi-infinite plate or wing which corresponds to $\varepsilon\rightarrow 0$.
Through numerical simulations, it was shown that \citeauthor{Pullin1978}'s asymptotic analysis can be extended to vortex rings where $\varepsilon=1$ \citep{Nitsche1994}.
The self-similar roll-up of the vortex sheet seems to be minorly affected by the geometric constraint as most of the vorticity in the core is concentrated in a viscous sub-core that is much smaller than the main vortex core \citep{Pullin1979}.

A non-dimensional Green function $\mathcal{G}_{\varepsilon}$, depending on $\varepsilon$, is defined as a function of the polar coordinates $s$ and $\theta$:
\begin{equation}
G(x,r,x',r')=R \, \mathcal{G}_{\varepsilon}(s,\theta,s',\theta')\quad.
\end{equation}
We now define the non-dimensional integral $\kindex{\mathcal{I}}{$\varepsilon,p$}$:
\begin{equation}
\kindex{\mathcal{I}}{$\varepsilon,p$}=\int\displaylimits^1_0 \int\displaylimits^1_0 \int\displaylimits^{2 \pi}_0 \int\displaylimits^{2\pi}_0 \left(ss'\right)^{p-1} \mathcal{G}_{\varepsilon}(s,\theta,s',\theta') \dd{s} \dd{s'} \dd{\theta} \dd{\theta'}\quad,
\end{equation}
and use it to express the energy:
\begin{equation}\label{eq:E}
E=\pi \kindex{\omega}{0}^2 R^5 \,\kindex{\mathcal{I}}{$\varepsilon,p$} \quad .
\end{equation}
The non-dimensional energy is derived from \cref{eq:IG,eq:E}:
\begin{equation}
\ksuper{E}{*}=\dfrac{p^2}{2 \pi \sqrt{2 \pi} }\left(\dfrac{2}{\varepsilon^2}+\dfrac{p}{p+2}\right)^{\nicefrac{-1}{2}} \,\kindex{\mathcal{I}}{$\varepsilon,p$}	\quad .
\end{equation}
This relationship only depends on the relative core radius $\varepsilon$ and the parameter $p$, which is directly related to the velocity profile of the incoming flow.
The integral $\kindex{\mathcal{I}}{$\varepsilon,p$}$ is computed numerically.\\

\begin{figure}
	\includegraphics{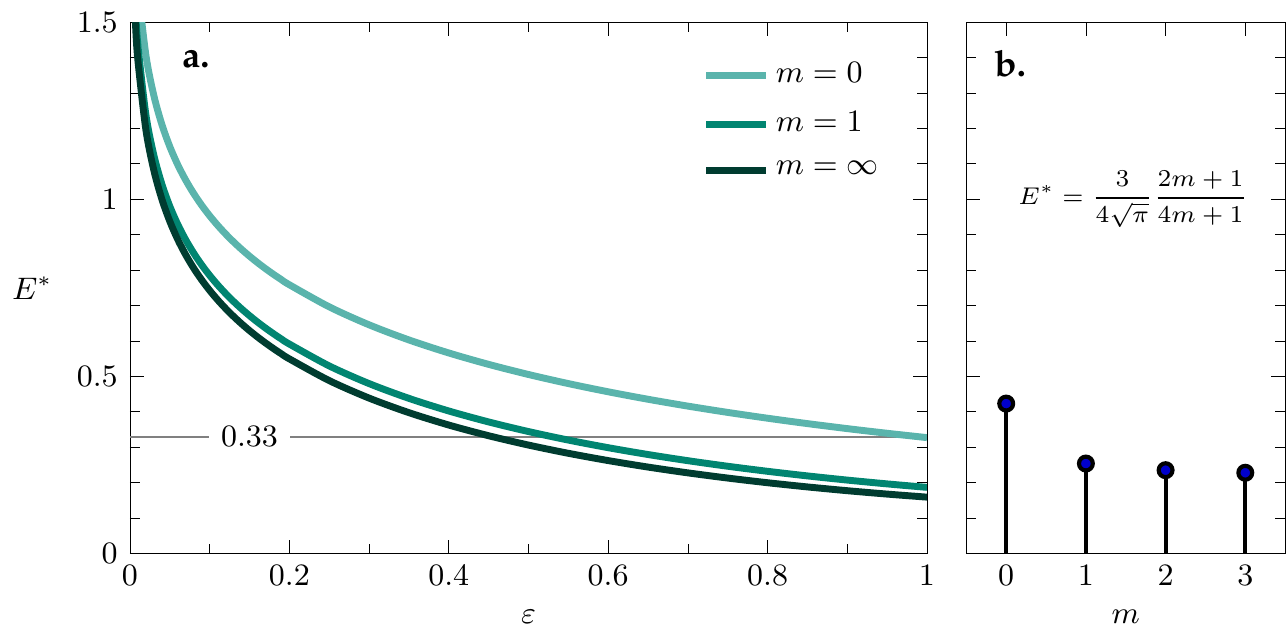}
	\caption[Pullin's non-dimensional energy]{\subf{a} Non-dimensional energy as a function of the relative core radius for different velocity profiles described by $u(t)\propto t^{m}$.
\subf{b} Non-dimensional energy as a function of $m$ according to \cref{shuss}.}
	\label{fig:Pull2}
\end{figure}

The evolution of the non-dimensional energy as a function of the relative core radius is presented in \cref{fig:Pull2}, for different flow velocity profiles.
With a relative core radius $\varepsilon=1$, the vorticity spreads to the axis of symmetry.
Yet, for low values of $m$, the majority of the vorticity remains concentrated near the centre of the vortex, in a viscous sub-core region that is an order of magnitude smaller than the overall core radius.
This is a characteristic feature of most experimentally observed vorticity distributions, which are often fitted with Gaussian profiles \citep{Weigand1997}.
For larger values of $m$, the distribution becomes smoother and ultimately reaches uniformity for $m\rightarrow\infty$.
Such uniform distributions are nearly impossible to produce experimentally, but they can be compared to theoretical distributions such as those corresponding to the \citeauthor{Norbury1973} family of vortices.

A vortex ring behind an axisymmetric object cannot grow beyond $\varepsilon=1$ and the minimum non-dimensional energy  $\ksubsuper{E}{$\varepsilon=1$}{*}$ is reached at this limit.
For a constant velocity ($m=0$), we compute $\ksubsuper{E}{$\varepsilon=1$}{*}=0.328$.
Experimentally, $\ksuper{E}{*}$ converges to approximately $0.3$ when the vorticity spreads to the axis of symmetry for ring vortices created by constant velocity jets or behind cones translating with constant velocity \citep{Gharib1998,DeGuyon2021}.
For a constant acceleration ($m=1$), $\ksubsuper{E}{$\varepsilon=1$}{*}=0.187$, which is in the range of the values measured by \citet{Dabiri2005a} and \citet{Mohseni2001}.
When $m$ tends to infinity, the vorticity distribution becomes uniform and $\ksubsuper{E}{$\varepsilon=1$}{*}=0.16$, which is the non-dimensional energy associated with Hill's spherical vortex.
For comparison, using the slug-flow model \pcref{shuss}, $\ksuper{E}{*}=0.21$ when $m\rightarrow\infty$.
The non-dimensional energy obtained with \cref{shuss} overestimates $\ksuper{E}{*}$ by approximately \SI{30}{\percent} for all values of $m$, due to inaccurate assumption on the shear layer velocity.
Both the slug-flow model and our Pullin's spiral based model predict a value of $\ksuper{E}{*}$ for $m=0$ that is twice the value of $\ksuper{E}{*}$ for $m=\infty$.
This is an interesting result given that both models have a different approach.
The presented model thus provides an accurate approximation of the non-dimensional energy for both propulsive and drag vortices regardless of any assumption on the shear layer velocity \citep{Limbourg2021c}.

\section{Diffusion of the Pullin's distribution}
\begin{figure}
\includegraphics[trim=0.5cm 0 0 0]{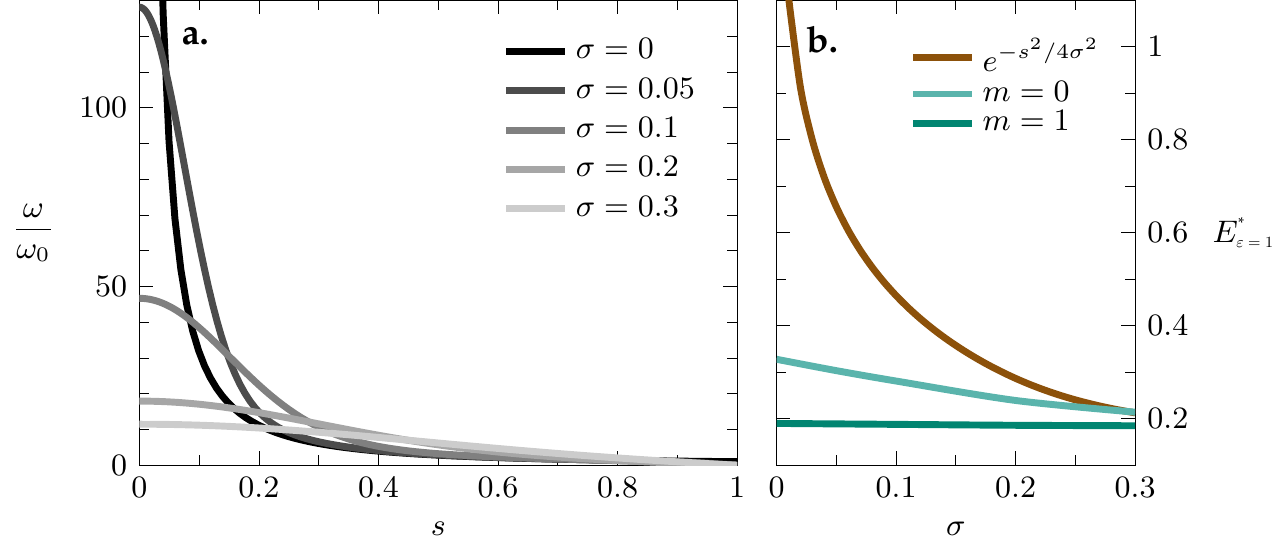}
\caption[Vorticity diffusion]{\subf{a} Vorticity distribution in the vortex core for $m=0$ and for different diffusion times $\sigma$.
\subf{b} The evolution of the minimum non-dimensional energy for $m=0$ and $m=1$ as a function of the viscous core size $\sigma$, compared with the minimum non-dimensional energy of a Lamb-Oseen vortex of similar core size.}
\label{fig:Pull3}
\end{figure}

The vorticity distribution that we derived in the previous section is valid for inviscid flows and is not representative of true vorticity distributions.
The inviscid vorticity distribution $\omega(s)=\kindex{\omega}{0} s^{p-2}$ \pcref{eq:Pull2}, has a singularity at the vortex centre $s=0$ as the exponent $p-2$ is negative \pcref{eq:Pull1}.
The vorticity is also discontinuous in $s=1$ by construction \pcref{eq:Pull2}.
To estimate the influence of viscosity, we propose here to radially diffuse the vorticity.
The vorticity is highly concentrated near the vortex centre in a viscous sub-core that is an order of magnitude smaller than the full core size \citep{Pullin1979}.
As a consequence, diffusion dominates convection in the sub-core for a vortex ring with $\varepsilon=1$ and the vorticity transport equation can be approximated by the two-dimensional equation:
\begin{equation}
\dfrac{\partial \omega}{\partial t}=\dfrac{\nu}{R^2} \dfrac{1}{s} \dfrac{\partial}{\partial s}\left(s \dfrac{\partial \omega}{\partial s} \right)
\end{equation}
with the boundary conditions:
\begin{align}
\dfrac{\partial \omega}{\partial s}\Bigr|_{\substack{s=0}}&=0\\[1ex]
\omega(s=1)&=0\quad.
\end{align}
The diffusion is integrated numerically during a non-dimensional diffusion time $\sigma=\sqrt{\nu t}/R$ with $\nu$ the kinematic viscosity of the fluid and $t$ the time.
Note that we are not considering the initial growth of the vortex ring, which is primarily dominated by convective time scales \citep{DeGuyon2021}.
Instead, we consider only the vorticity distributions corresponding to $\varepsilon=1$, as it is the limiting value of the vortex formation process.
The non-dimensional quantity $\sigma$ should not be interpreted as a non-dimensional diffusion time but rather as the relative size of the viscous sub-core with respect to the ring size.
It is a measure for the centre region that contains the majority of the vorticity distributed within the vortex core.
The diffused vorticity profiles for different values of $\sigma$ and the corresponding evolution of the non-dimensional energy are presented in \cref{fig:Pull3}.
Overall, the maximum vorticity in the vortex centre decreases and the vorticity distribution widens with increasing values of the non-dimensional diffusion time (\cref{fig:Pull3}a).
The broader, more uniform vorticity distributions at higher values of $\sigma$ correspond to lower values of the minimum non-dimensional energy (\cref{fig:Pull3}b).
For a constant velocity ($m=0$), the non-dimensional energy drops from $\ksubsuper{E}{$\varepsilon=1$}{*}=0.328$ to $\ksubsuper{E}{$\varepsilon=1$}{*}=0.281$ when the viscous core size is increased from $\sigma=0$ to $\sigma=0.1$.
These minimum values of the non-dimensional energy remain in agreement with the values measured by \citet{Gharib1998,Mohseni2001,DeGuyon2021}.
In the case of a constant acceleration motion ($m=1$), the inviscid vorticity profile already has a low non-dimensional energy of $0.187$ for $\sigma=0$, which is close to the limiting value of $0.16$ obtained with a uniform vorticity distribution.
The diffusion during a non-dimensional time of $\sigma=0.3$ leads to a decrease of  $\ksubsuper{E}{$\varepsilon=1$}{*}$ of less than \SI{5}{\percent} for motions with a constant acceleration.

\begin{figure}
	\includegraphics{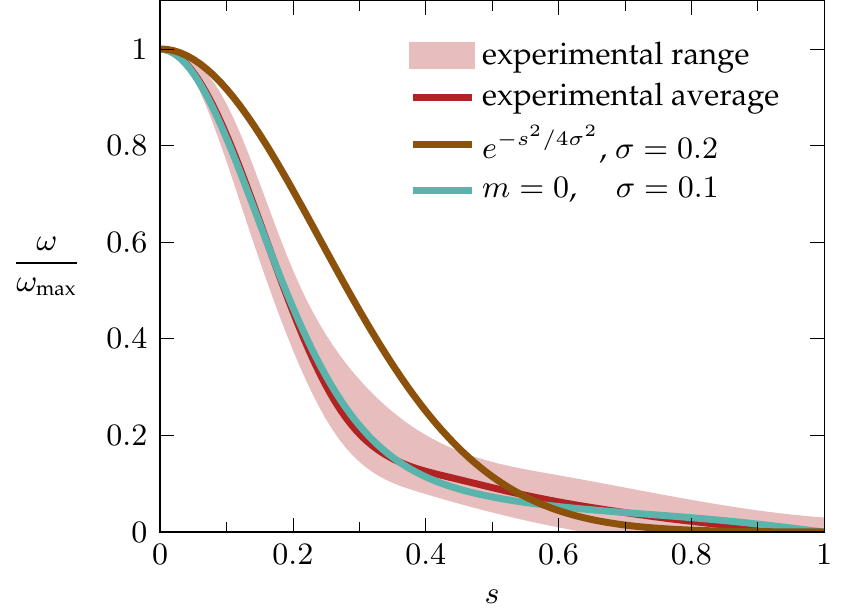}
	\caption[Vorticity profile]{Vorticity distribution for:
	the diffused Pullin profile for $m=0$ and $\sigma=0.1$,
	the experimental vorticity profiles from \cite{DeGuyon2021}, measured between the vortex core and the $x$-axis of symmetry,
	and the Lamb-Oseen profile of equivalent non-dimensional energy ($\sigma=0.2$).
	The experimental data is presented by the average vorticity profile (solid line) and its range (shaded region).}
	\label{fig:Pull4}
\end{figure}

We now compare the results of the Pullin distribution with a time varying Lamb-Oseen distribution:
\begin{equation}
\omega(s)=\kindex{\omega}{0} e^{-(sR)^2/4\nu t}\quad,
\end{equation}
where $\kindex{\omega}{0}$ is again the reference vorticity.
Using $\sigma=\sqrt{\nu t}/R$, this is equivalent to:
\begin{equation}
\omega(s)=\begin{cases}
\kindex{\omega}{0} e^{-s^2/4\sigma^2} ,& \text{if } s\leq 1\\
0,              & \text{if } s> 1\quad.
\end{cases}
\label{eq:lamb}
\end{equation}
The Lamb-Oseen distribution is often used to fit experimental or numerical vortex core distributions \citep{Weigand1997,Johari2002}.
It corresponds to the diffusion of a point vortex or a highly concentrated vorticity distribution.
Therefore, the Lamb-Oseen vortex has a higher non-dimensional energy than the diffused rolled-up Pullin spiral (\cref{fig:Pull3}b).
The differences between the Pullin and the Lamb-Oseen distribution vanish when $\sigma$ increases, and for $\sigma=0.3$ both solutions have a similar non-dimensional energy.
A Pullin vortex with $\sigma=0.1$ and a Lamb-Oseen vortex with $\sigma=0.2$ produce an equivalent $\ksubsuper{E}{$\varepsilon=1$}{*}=0.28$.
Their vorticity distribution is shown in \cref{fig:Pull4}, and compared to the experimentally measured vorticity profiles of a vortex ring forming behind a disk, presented by \citet{DeGuyon2021}.
The diffused Pullin profile matches the experimental profiles better than the Gaussian Lamb-Oseen profile, which has a core that is too wide and a tail that is too low compared to the experimental results.
A better fit of the core region can be achieved with a Lamb-Oseen distribution of parameter $\sigma=0.12$, but this distribution drops too quickly to zero and yields a value of the non-dimensional energy that is substantially overestimated.

\section{Conclusion}
The non-dimensional energy of vortex rings produced by the impulsive start of a piston converges to \num{0.33}.
Values around \num{0.3} were also obtained for various experiments and simulations, including for vortex rings generated behind disks.
Here, we explored the limits of the universality of this value and analysed the variations that occur outside of those limits based on the self-similar vortex sheet roll-up described by \citeauthor{Pullin1978}.
We derived the vorticity distribution that emerges when an inviscid shear layer rolls up into a spiral.
Even though the vorticity distribution is singular, this approach yields a finite non-dimensional energy of \num{0.33} for a piston moving at a constant velocity.

The distribution of vorticity is strongly influenced by the kinematics of the vortex generator.
Switching from a constant velocity profile to a linearly increasing velocity profile reduces the non-dimensional energy to \num{0.19}.
This behaviour was observed for vortex rings generated by piston-cylinders and results in delayed vortex pinch-off.
This offers opportunities to control the vortex development and improve fluid transport or thrust production.
Further increasing the acceleration, i.e. increasing $m$ in the velocity program $u(t) \propto t^m$, only reduces $\ksuper{E}{*}$ down to \num{0.16}.
In addition to being unpractical to achieve, increasing the acceleration does not yield much improvement in the tuning of the vortex characteristics.

The influence of viscosity on the limiting minimum value of the non-dimensional energy is analysed by radially diffusing the vorticity in the vortex core.
The radial diffusion of vorticity from the rolled-up shear layer removes the singularity in the distribution and yields vorticity profiles that, compared to a traditional Gaussian distribution, describe more accurately the profiles measured in previous work.
The diffused Pullin profile predicts a value of the non-dimensional energy of \num{0.28} in accordance with the experimental observations.

The proposed model offers a practical alternative to the existing slug-flow model.
It is more broadly applicable than the slug-flow model and requires only the kinematics of the vortex generator as input.

\subsection*{Declaration of interests.}
The authors report no conflict of interest.
\bibliographystyle{jfm}
\bibliography{p3}
\end{document}